\documentstyle[12pt,epsfig]{article}
\begin{document}
\begin{flushright}
{\bf Preprint SSU-HEP-01/02\\
Samara State University}
\end{flushright}
\vspace{30mm}
\centerline{\bf NUCLEUS POLARIZABILITY CONTRIBUTION}
\baselineskip=13pt
\centerline{\bf TO THE HYDROGEN-DEUTERIUM ISOTOPE SHIFT}
\vspace*{0.37truein}
\centerline{\footnotesize R.N.FAUSTOV\footnote{E-mail:
faustov@theory.npi.msu.su}}
\baselineskip=12pt
\centerline{\footnotesize\it Scientific Council for Cybernetics RAS,
Vavilov 40}
\baselineskip=10pt
\centerline{\footnotesize\it Moscow, 117333, Russian Federation}
\vspace*{10pt}
\centerline{\footnotesize A.P.MARTYNENKO\footnote{E-mail:
mart@info.ssu.samara.ru}}
\baselineskip=12pt
\centerline{\footnotesize\it Samara State University, Department of
Theoretical Physics, Pavlov 1}
\baselineskip=10pt
\centerline{\footnotesize\it Samara, 443011, Russian Federation}
\vspace*{0.225truein}

\vspace*{0.21truein}
\begin{abstract}
The correction to the hydrogen-deuterium isotope shift
due to the proton and deuteron polarizability is evaluated on the
basis of modern experimental data on the structure functions of
inelastic lepton-nucleus scattering. The numerical value of this
contribution is equal $63\pm 12$ Hz.
\end{abstract}

\vspace*{10pt}
{\rm keywords: Hydrogen atom, Lamb shift,
Nucleus Polarizability}

\newpage

The study of the hydrogen atom energy levels, which is carried out during many years,
open the possibility for testing Quantum Electrodynamics with
unprecedented high accuracy \cite{EGS}. As the simplest stable
atom it has essential experimental advantage in the comparison
with the other hydrogenic atoms. One of the remarkable recent
experiments is the measurement of the isotope
shift in atomic hydrogen \cite{Huber}. The value of the
difference of the mean square charge radii of the proton and
deuteron was obtained from this experiment. The measurement \cite{Huber}
of the hydrogen-deuterium isotope shift in the 1S-2S
level splitting
\begin{equation}
\Delta E_{H-D}(1S-2S)=670~994~334.64(15)~kHz.
\end{equation}
The experimental accuracy in this case is too high that there is need to take
into account the effects of the strong interaction for succesful comparison
of the theory and experiment and extraction the difference of the sizes
of proton and deuteron. The influence of strong interaction on the
structure of the hydrogen-like atoms energy levels is manifested in some
effects: the hadronic vacuum polarization, the nucleus structure corrections,
the nucleus polarizability corrections \cite{FM1,FM2}. The electron polarizes
the proton or the deuteron and produces nuclear polarizability contribution
to the isotope shift (1). Nuclear polarizability corrections in $\Delta E_{H-D}$
were studied in \cite{Friar,KS}, where the spin independent Compton forward
scattering amplitude was expressed through the electric polarizability $\alpha$
and magnetic polarizability $\beta$. In this case the nuclear polarizability
contribution was calculated in the logarithmic approximation and the term,
connected with the neutron magnetic polarizability was omitted due to the
absence of the reliable experimental data on $\beta_n$. At the same time
it is known that the $\gamma$-deuteron Compton scattering probes the structure
of the deuteron and provides necessary information that may allow the extraction
of neutron properties, such as neutron electric and magnetic polarizabilities
\cite{NMC1,NMC2,Hornidge,Kolb,LL}.
The higher-order precision approach to the nucleus polarizability calculation
was suggested in \cite{FM1}. This method is based on the using of the
experimental data about inelastic lepton-nucleus scattering structure functions.
The imaginary part of the forward Compton scattering amplitude of the virtual
photon is determined by the tensor
\begin{equation}
W^{\mu\nu}(p,q)=F_1(x,Q^2)\left(-g^{\mu\nu}-\frac{q^\mu q^\nu}{q^2}\right)
+\frac{F_2(x,Q^2)}{(p\cdot q)}\left(p^\mu-\frac{p\cdot q}{q^2}q^\mu\right)
\left(p^\nu-\frac{p\cdot q}{q^2}q^\nu\right),
\end{equation}
where $q^2=-Q^2$ is the square of the four momentum transfer, $x=Q^2/2pq$ is
the Bjorken scaling variable. In the nucleus rest frame $p\cdot q=m_2\nu$,
$m_2$ is the nucleus mass. The invariant mass squared of the electroproduced
hadronic system $W^2=m_2^2+2m_2\nu-Q^2$. The tensor $W^{\mu\nu}$ determines
the nucleus polarizability contribution to the Lamb shift in the hydrogenic
atom, which can be written in the form \cite{FM1,P}:
\begin{equation}
\Delta
E_{pol}^{Ls}=-\frac{16\mu^3(Z\alpha)^5m_1}{\pi^2n^3}\int_0^\infty\frac{dk}{k}
\int_0^{\pi}d\phi\int_{\nu_0}^\infty
dy\frac{\sin^2\phi}{(k^2+4m_1^2\cos^2\phi)
(y^2+k^2\cos^2\phi)}\times
\end{equation}
\begin{displaymath}
[(1+2\cos^2\phi)\frac{(1+k^2/y^2)\cos^2\phi}{1+R(y,k^2)}+
\sin^2\phi]F_2(y,k^2)+\frac{2\mu^3\alpha^5}{\pi
n^3m_1m_2}\int_0^\infty h(k^2)\beta(k^2) k dk,
\end{displaymath}
where $R=\sigma_L/\sigma_T$ is the ratio of the longitudinally to transversely
polarized virtual photon absorption cross sections, $\nu_0$ is the
threshold value of the transfer energy $\nu$ for production of $\pi$-mesons,
\begin{equation}
h(k^2)=1+\left(1-\frac{k^2}{2m_1^2}\right)\left(\sqrt{\frac{4m_1^2}{k^2}+1}-1\right).
\end{equation}
The second part of the eq. (3)
is connected with the subtraction term in the dispersion relation for the
Compton scattering amplitude. The dipole parameterization for it was proposed
in \cite{P}:
\begin{equation}
\beta(k^2)=\beta \cdot
G(k^2),~~G(k^2)=\frac{1}{\left(1+k^2/0.71\right)^2},
\end{equation}
where $\beta$ is the nucleus magnetic polarizability.
It is necessary to emphasize that the expression (3) is valid
both for the hydrogen and deuterium, which has the spin 1
nucleus. In the region $Q^2>0.5~Gev^2$ there is 6-parameter model
for the quantity $R(x,Q^2)$ \cite{W}, which was used in the
hydrogen Lamb shift calculation \cite{FM1}. But the most
important contribution to (3) comes from the resonance region with
$Q^2\approx 0$. There are predictions by several models that
$R\sim (Q^2)^{c_1}$ at small $Q^2$ \cite{ZEUS}. So, we considered
$R\approx 0$ in this region. All experimental investigations show
also that $R_p\approx R_d$ with high precision \cite{NMC2}.

The calculation of the proton polarizability contribution to the Lamb shift
of the hydrogen atom
was carried out in \cite{FM1} by means of (3). The deuteron structure
function $F_2^d$ was derived on the basis of numerous experimental data
on deep inelastic scattering. The complete parameterization of $F_2^d$
including the resonance region was obtained in the form \cite{NMC1}:
\begin{equation}
F_2^d=\left[1-G^2(Q^2)\right]\left[F^{dis}(x,Q^2)+F^{res}(x,Q^2)+F^{bg}(x,Q^2)\right],
\end{equation}
where the contribution from deep inelastic region was parameterized as
\begin{equation}
F^{dis}(x,Q^2)=\left[\frac{5}{18}\frac{3}{B(\eta_1,\eta_2+1)}x^{\eta_1}_w
(1-x_w)^{\eta_2}+\frac{1}{3}\eta_3(1-x_w)^{\eta_4}\right]S(x,Q^2),
\end{equation}
\begin{displaymath}
x_w=\frac{Q^2+m_a^2}{2m_2\nu+m_b^2},~~~S(x,Q^2)=1-e^{-a(W-W_{thr})},
\end{displaymath}
the contribution from the resonance region
\begin{equation}
F^{res}(x,Q^2)=\alpha_5^2G^{3/2}e^{-(W-m_\Delta)^2/\Gamma^2},
\end{equation}
and the background under the resonance region was parameterized as follows:
\begin{equation}
F^{bg}(x,Q^2)=\alpha_6^2G^{1/2}\xi e^{-b(W-W_{thr})^2}.
\end{equation}
The values of all parameters in (7)-(9) were taken from \cite{NMC1}. The
magnetic polarizability of deuteron $\beta_d$ determines the contribution
of the second term in relation (3). We considered, that $\beta_d\approx \beta_p+
\beta_n$. The proton magnetic polarizability \cite{RPP}
\begin{equation}
\beta_p=2.1\pm 0.8\pm 0.5 \cdot 10^{-4}~fm^3.
\end{equation}
There are contradictory data on the neutron magnetic polarizability at present
\cite{Hornidge,Kolb,LL}, so we proposed that $\beta_p\approx\beta_n$. Taking
into account the parameterization of the proton and the deuteron structure
functions we performed numerical integration in (3). As a result, the proton
polarizability contribution to 1S-2S interval
\begin{equation}
\Delta E_H(1S-2S)=-68\pm 9~Hz,
\end{equation}
and in the case of deuterium the analogous quantity
\begin{equation}
\Delta E_D(1S-2S)=-131\pm 8~Hz.
\end{equation}
The theoretical uncertainties in eq. (11) and (12) are determined
by the experimental errors when measuring the structure function
$F_2$ and the proton magnetic polarizability (10). The
corresponding polarizability contribution to the isotope shift
between hydrogen and deuterium 1S-2S transition can be presented
as
\begin{equation}
\Delta E_{H-D}(1S-2S)=63\pm 12~Hz.
\end{equation}
The errors of (11) and (12) were added in quadratures.

As was pointed out above, the expression for the electron-nucleus
interaction operator taking into account the polarizability
effects was obtained \cite{Friar,KS} in the coordinate
representation in terms of the electric polarizability
$\bar\alpha(0)$: $V=-5\alpha m_e\bar\alpha(0)\cdot$ $\cdot\ln(\bar
E/m_e)\delta(\vec r)$. The corresponding numerical value of the
contribution to the H-D isotope shift due to the internal
polarizabilities of the nucleons equals $53\pm 9\pm
11$ Hz \cite{KS}. So, our numerical value (13) is in good
agreement with the results of \cite{Friar,KS}.

The obtained result (13) is comparable with the experimental
accuracy 150 Hz from \cite{Huber}. Our method for the study of
nucleus polarizability contribution, based on eq. (3), provides a
way to increase the accuracy of the calculation considering more
exactly the experimental data on the structure functions
$F_2(x,Q^2)$ and $R(x,Q^2)$. The charge radius of the proton
$r_p$ (or deuteron $r_d$) is one of the universal fundamental
physical constants, because it is important in many different
physical tasks \cite{SGK}. More detailed experimental study of
the energy levels in the hydrogenic atoms can improve our
knowledge of $r_p$ and $r_d$. The measurement of 2P-2S Lamb shift
in muonic hydrogen with an error 8 $\mu eV$ \cite{FK} will allow
to determine the proton charge radius with relative accuracy about
$0.1\%$. Similar accuracy for the deuteron charge radius can be
reached thereafter from the measurement of the H-D isotope shift
with the precision 50 Hz after improving the
precision of the electron-proton mass ratio measurement \cite{EGS}.\\[3mm]
{\bf Acknowledgements}

We are grateful to
I.B.Khriplovich, R.A.Sen'kov for useful discussions. The work was
performed under the financial support of the Russian Foundation
for Fundamental Research (grant 00-02-17771), the Program
"Universities of Russia" (grant 990192) and the Ministry of Education
(grant E00-3.3-45).


\begin{thebibliography}{99}
\bibitem{EGS} M.I. Eides, H. Grotch and V.A. Shelyuto, Phys. Rep.
{\bf 342}, 62 (2001).
\bibitem{Huber} A. Huber, Th. Udem, B. Gross et al.,  Phys. Rev. Lett.
{\bf 80}, 468 (1998).
\bibitem{FM1}R.N. Faustov, A.P. Martynenko, Phys. Atom. Nuclei
{\bf 63}, 915 (2000).
\bibitem{FM2}
R.N. Faustov, A. Karimkhodzhaev, A.P. Martynenko, Phys. Rev.
{\bf A59}, 2498 (1999).
\bibitem{Friar}
J.L. Friar, G.L. Payne, Phys. Rev.
{\bf C55}, 2764 (1997).
\bibitem{KS}
I.B. Khriplovich, R.A. Sen'kov, Phys. Lett.
{\bf A249}, 474 (1998).
\bibitem{NMC1}
P. Amaudruz, M. Arneodo, A. Arvidson, et al., Nucl. Phys.
{\bf B273}, 3 (1992).
\bibitem{NMC2}
M. Arneodo, A. Arvidson, B. Badelek, et al., Nucl. Phys.
{\bf B487}, 3 (1997).
\bibitem{Hornidge}
D.L. Hornidge, B.J. Warkentin, R. Igarashi, et al., Phys. Rev. Lett.
{\bf 84}, 2334 (2000).
\bibitem{Kolb}
N.R. Kolb, A.W. Rauf, R. Igarashi, et al., Phys. Rev. Lett.
{\bf 85}, 1388 (2000).
\bibitem{LL}
M.I. Levchuk, A.I. L'vov, E-preprint nucl-th {\bf /0010059}, (2000).
\bibitem{P}
K. Pachucki, Phys. Rev.
{\bf A60}, 3593 (1999).
\bibitem{W}
K. Abe, et al., Preprint SLAC-PUB-{\bf 7927}, (1998).
\bibitem{ZEUS}
K. Ackerstaff, et al., Preprint DESY {\bf 99-199}, (2000).
\bibitem{RPP}
Review of Particle Physics, Eur. Phys. Jour.
{\bf C15}, 1 (2000).
\bibitem{SGK}
S.G. Karshenboim, in Proc. Int. Workshop "Hadronic atoms and positronium
in the standard model", ed. M.A. Ivanov, et al., (Dubna, 1998), 224.
\bibitem{FK}
F. Kottmann, in Proc. QED 2000 2nd Workshop on Frontier Tests of Quantum
Electrodynamics and Physics of the Vacuum, ed. D. Amati, et al.,
(Trieste, 2000).
\end{thebibliography}
\end{document}